\documentclass[twocolumn,floatfix,preprintnumbers,citeautoscript,newabstract]{revtex4} %,floats
\usepackage{amsmath,amssymb,graphics}
\usepackage{bbm} 
\newcommand{\CenterObject}[1]{\ensuremath{\vcenter{\hbox{#1}}}}
\newcommand{\I}{\mathrm{i}}
\newcommand{\E}[1]{\ensuremath{\mathrm{E}_{#1}}} % e.g. \E{8}
\newcommand{\G}[1]{\ensuremath{\mathrm{G}_{#1}}}
\newcommand{\SO}[1]{\ensuremath{\mathrm{SO}(#1)}}
\newcommand{\SU}[1]{\ensuremath{\mathrm{SU}(#1)}}
\newcommand{\U}[1]{\ensuremath{\mathrm{U}(#1)}}
\newcommand{\Z}[1]{\ensuremath{\mathbbm{Z}_{#1}}} % Z_N ->\Z{N}

\def\pl#1#2#3{Phys.~Lett.~B~{\bf {#1}} ({#2}) #3}
\def\np#1#2#3{Nucl.~Phys.~B~{\bf {#1}} ({#2}) #3}

\begin{document} \setlength{\unitlength}{1in}

\preprint{DESY 05-222}
    
\title{Supersymmetric Standard Model from the Heterotic String}

\author{Wilfried~Buchm\"uller$^1$, Koichi Hamaguchi$^1$, 
Oleg Lebdedev$^2$, Michael Ratz$^2$}

\affiliation{%
${}^1$Deutsches Elektronen-Synchrotron DESY, 22603 Hamburg, Germany,\\
$^2$Physikalisches Institut der Universit\"at Bonn, Nussallee 12, 53115 Bonn, Germany}

\begin{abstract}
We present a $\Z6$ orbifold compactification of the  $\E8\times\E8$  heterotic
string which leads to the  (supersymmetric)    Standard Model gauge group and
matter content. The quarks and leptons  appear as three 
$\boldsymbol{16}$--plets of $\SO{10}$, whereas the Higgs fields do not form
complete  $\SO{10}$ multiplets. The model has large vacuum degeneracy. For
generic vacua, no exotic states appear at low energies and the model is
consistent with gauge coupling unification. The top quark Yukawa coupling arises
from gauge interactions and is   of the order of  the gauge couplings, whereas
the other Yukawa couplings are suppressed. 
\end{abstract}

\pacs{\dots}

\maketitle

The problem of ultraviolet completion of the (supersymmetric) Standard Model
(SM) has been a long--standing issue in particle physics. The most promising
approach is based on string theory, however explicit models usually contain
exotic particles and suffer from other phenomenological problems. The purpose of
this Letter is to show that these difficulties can be overcome in the well known
weakly coupled heterotic string \cite{ghx85} compactified on an orbifold
\cite{dhx85,inx87}. The emerging picture has a simple geometrical
interpretation.  

In the light cone gauge the heterotic string can be described by
the following bosonic world sheet fields: 8 string coordinates $X^i$, $i=
1\ldots 8$, 16 internal left-moving coordinates $X^I$, $I=1\ldots 16$, and 4
right-moving fields $\phi^i$, $i=1\ldots 4$, which correspond to the bosonized
Neveu-Schwarz-Ramond fermions (cf.~\cite{kkx90}). The 16 left-moving
internal coordinates are compactified on a torus. The associated quantized
momenta lie on the $\E8\times \E8$ root lattice. 
The massless spectrum of this 10D string is 10D supergravity coupled to 
$\E8\times \E8$ super Yang--Mills theory.

To get an effective four--dimensional theory, 6 dimensions of the 10D heterotic
string are compactified on an orbifold. A $\Z{N}$ orbifold is obtained by
modding a 6D torus together with the 16D gauge torus by a $\Z{N}$ twist, ${\cal
O}\ =\ \mathbbm{T}^6 \otimes \mathbbm{T}_{\E8\times \E8'} / \Z{N}$. On the three
complex torus coordinates $z^i$, $i=1,2,3$, the $\Z{N}$ twist acts as $z^i
\rightarrow e^{2\pi \I\, v_N^i}~ z^i $, where $Nv_N$ has integer components and
$\sum_i v_N^i =0$. This action is accompanied by the shifts of the bosonized
fermions $\phi^i$ and the gauge coordinates $X^I$, $\phi^i \rightarrow \phi^i -
\pi v_N^i\;, \quad X^I \rightarrow X^I + \pi V_N^I $, where $NV_N$ is an
$\E8\times \E8$ lattice vector. Further, if discrete Wilson lines $W^I_l$ are
present \cite{inx87}, the torus lattice translations are accompanied by the
shifts $X^I \rightarrow X^I + \pi n_l W^I_l $ with integer $n_l$.  In addition
to the requirement that $NV_N$ and $nW$, where $n$ is the order of the Wilson
line, be $\E8\times\E8$ lattice vectors, $V_N$ and $W$ are constrained by
modular invariance (cf.\ \cite{Kobayashi:2004ud,Forste:2004ie}). At each fixed
point of the orbifold, a local \Z{N} twist, composed of $V_N$ and discrete
Wilson lines, breaks $\E8\times\E8$ to a subgroup (cf.\
\cite{Buchmuller:2004hv}). Given an orbifold, a torus lattice together with the
shift $V_N$ and the Wilson lines, the massless spectrum of the orbifold can be
calculated. It is supersymmetric by construction and consists of the states
which are invariant under the twisting and lattice translations. 

%%%%%%%%%%%%%%%%%%%%%%%%%%%%%%%%%%%%%%%%%%%%%%%%%%%%%%%%%%%%
\begin{figure}[h]
\centerline{\CenterObject{\includegraphics{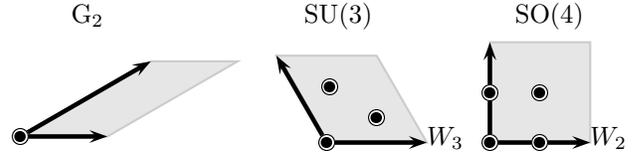}}
}
\caption{$\G2\times\SU3\times\SO4$ torus lattice of a $\Z6$-II orbifold.}
\label{fig1}
\end{figure}
%%%%%%%%%%%%%%%%%%%%%%%%%%%%%%%%%%%%%%%%%%%%%%%%%%%%%%%%%%%%
In our construction we choose a $\Z6$-II orbifold based on a Lie torus lattice
$\G2\times\SU3\times\SO4$ (Fig.~\ref{fig1})  with a twist vector
$v_6=\left(\frac{1}{6}, \frac{1}{3}, -\frac{1}{2} \right)$
\cite{Kobayashi:2004ud}. In addition to $V_6$, we require two Wilson
lines: one of  order two in the $\SO4$ plane, and another of order three in
the $\SU3$ plane. In an orthonormal basis, the shift and the Wilson lines  are
given by
\begin{eqnarray}
 V_6 & = &
 \left(\tfrac{1}{2},\tfrac{1}{2},\tfrac{1}{3},0,0,0,0,0\right) \, 
 \left(\tfrac{1}{3},0,0,0,0,0,0,0\right) 
 \;,\nonumber\\
 W_2 & = & 
 \left(\tfrac{1}{2},0,\tfrac{1}{2},\tfrac{1}{2},\tfrac{1}{2},0,0,0\right) 
 \,\left(-\tfrac{3}{4},\tfrac{1}{4},\tfrac{1}{4},-\tfrac{1}{4},\tfrac{1}{4},\tfrac{1}{4},\tfrac{1}{4},-\tfrac{1}{4}\right) 
 \;,\nonumber\\
 W_3 & = &
 \left(\tfrac{1}{3},0,0,\tfrac{1}{3},\tfrac{1}{3},\tfrac{1}{3},\tfrac{1}{3},\tfrac{1}{3}\right) \, 
 \left(1,\tfrac{1}{3},\tfrac{1}{3},\tfrac{1}{3},0,0,0,0\right)
 \;. \nonumber
\end{eqnarray}
The model has twelve fixed points which come in six inequivalent pairs, with the
local groups
\[\begin{array}{l}
 \SO{10}\times\SO{4}\;,\;\SO{12}\;,\;\SU7\;,\;\\
 \SO{8}\times\SO6\;,\;\SO{8}'\times\SO6'\;,\;\SO{8}''\times\SO6''
\end{array}\]
up to \U1 factors and subgroups of the second \E8. The standard model gauge
group, $G_\mathrm{SM}=\SU3_c\times\SU2_\mathrm{L}\times\U1_Y$, is obtained as an
intersection of those groups. The surviving gauge group in 4D is 
\begin{equation}
 G~=~G_\mathrm{SM}\times\left[\SO{6}\times\SU2\right]\times\U1^8\;,
\label{G}
\end{equation}
where one of the $\U1$'s is anomalous, and the brackets indicate a subgroup of
the second \E8.  The matter multiplets are found by solving the masslessness
equations together with the twist- and translation--invariance conditions.  The
resulting spectrum includes both untwisted  ($U$, ``bulk'') and twisted ($T_k$,
``localized'') states,  and is given in Tab.~\ref{tab:Spectrum}. The twisted
states can belong to any of the twisted sectors $T_k$ ($k=1,2,3,4$) depending on
their string boundary conditions. There are no left--chiral  superfields in
the $T_5$ sector.
%%%%%%%%%%%%%%%%%%%%%%%%%%%%%%%%%%%%%%%%%%%%%%%%%%
\begin{table}[!t]
\centerline{
 \begin{tabular}{|c|c|c|c|c|c|c|}
 \hline
 name & Representation & count & &
 name & Representation & count\\
 \hline
 $q_i$ & $(\boldsymbol{3},\boldsymbol{2};\boldsymbol{1},\boldsymbol{1})_{1/6}$ & 3 & &
 $\bar u_i$ & $(\overline{\boldsymbol{3}},\boldsymbol{1};\boldsymbol{1},\boldsymbol{1})_{-2/3}$ 
 	& 3\\
 $\bar d_i$ & $(\overline{\boldsymbol{3}},\boldsymbol{1};\boldsymbol{1},\boldsymbol{1})_{1/3}$ 
 	& 7 & & 
 $d_i$ & $(\boldsymbol{3},\boldsymbol{1};\boldsymbol{1},\boldsymbol{1})_{-1/3}$ & 4\\
 $\bar\ell_i$ & $(\boldsymbol{1},\boldsymbol{2};\boldsymbol{1},\boldsymbol{1})_{1/2}$ & 5 & &
 $\ell_i$ & $(\boldsymbol{1},\boldsymbol{2};\boldsymbol{1},\boldsymbol{1})_{-1/2}$ & 8\\
 $m_i$ & $(\boldsymbol{1},\boldsymbol{2};\boldsymbol{1},\boldsymbol{1})_{0}$ & 8 & &
 $\bar e_i$ & $(\boldsymbol{1},\boldsymbol{1};\boldsymbol{1},\boldsymbol{1})_{1}$ & 3 \\
 $s^-_i$ & $(\boldsymbol{1},\boldsymbol{1};\boldsymbol{1},\boldsymbol{1})_{-1/2}$ & 16 & & 
 $s^+_i$ & $(\boldsymbol{1},\boldsymbol{1};\boldsymbol{1},\boldsymbol{1})_{1/2}$ & 16\\
 $s^0_i$ & $(\boldsymbol{1},\boldsymbol{1};\boldsymbol{1},\boldsymbol{1})_{0}$ &
 69
 & & $h_i$ &
 $(\boldsymbol{1},\boldsymbol{1};\boldsymbol{1},\boldsymbol{2})_{0}$& 14\\
 $f_i$ & $(\boldsymbol{1},\boldsymbol{1};\boldsymbol{4},\boldsymbol{1})_{0}$ & 4 & &
 $\bar f_i$ &
 $(\boldsymbol{1},\boldsymbol{1};\overline{\boldsymbol{4}},\boldsymbol{1})_{0}$ & 4 \\
 $w_i$ & $(\boldsymbol{1},\boldsymbol{1};\boldsymbol{6},\boldsymbol{1})_{0}$ &
 5 & & & &\\
 \hline
 \end{tabular}	
}
\caption{The $G_\mathrm{SM}\times[\SO{6}\times\SU2]$ quantum numbers
of the spectrum. }
\label{tab:Spectrum} 
\end{table} 
%%%%%%%%%%%%%%%%%%%%%%%%%%%%%%%%%%%%%%%%%%%%%%%%%

\begin{figure}[t!]
\centerline{\CenterObject{\includegraphics{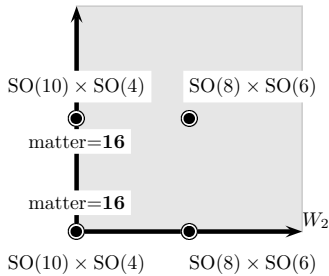}}
}
\caption{ Local gauge symmetries in the $\SO4$ plane (at the origin  in the
$\G2$ and $\SU3$ planes). Two $\boldsymbol{16}$--plets arise from the
orbifold fixed points. }
\label{fig2}
\end{figure}

Let us now discuss some properties of the spectrum. We first note that the $V_6$
shift is chosen such that the local gauge symmetry at the origin is
$\SO{10}\times\SO4\times\U1$ and the twisted matter at this point is a
$\boldsymbol{16}$--plet of $\SO{10}$ plus $\SO{10}$-singlets. When the
$\SU3\times\SU2\times\U1 \subset \SO{10}$ factor is identified with the 
SM group
$G_\mathrm{SM}$ (with the standard GUT hypercharge embedding), the $\boldsymbol{16}$--plet of $\SO{10}$ gives a complete
generation of the SM matter, including the right--handed neutrino. Since there
are two equivalent fixed points in the $\SO4$ plane (Fig.~\ref{fig2}), there are
2 copies of the $\boldsymbol{16}$--plets.  Due to our choice of Wilson lines,
the remaining matter has the SM quantum numbers of an additional
$\boldsymbol{16}$--plet plus vector--like multiplets. This can partly be
understood from the SM anomaly cancellation. Thus, we have
\begin{equation}  
 \text{matter:}~~~~3\times\boldsymbol{16} ~~+~~ \text{vector-like} \;.
\end{equation}  
Two generations are localized in the compactified space and come from the first
twisted sector $T_1$, whereas the third generation is partially twisted and partially
untwisted:
\begin{equation}
2\times\boldsymbol{16} \in T_1 ~~,~~ \boldsymbol{16} \in U,T_2,T_4 \;.
\end{equation}
In particular, the up--quark and the quark doublet of the third generation
are untwisted, which results in a large Yukawa coupling, whereas the down--quark
is twisted and its Yukawa coupling is suppressed.

%%%%%%%%%%%%%%%%%%%%%%%%%%%%%%%%%%%%%%%%%%%%%%%%%%%%%%%
%%%%%%%%%%%%%%%%%%%%%%%%%%%%%%%%%%%%%%%%%%%%%%%%%%%%%%%%% 

It is well known that the heterotic orbifold models have large 
vacuum degeneracy which gives enough freedom for realistic constructions
\cite{Casas:1988se,Font:1988mm,Font:1989aj}. 
There are many flat directions in the field
space along which supersymmetry is preserved but some of the gauge symmetries
are broken. In particular, all of the $\U1$ factors of Eq.~\eqref{G} apart from
the hypercharge are broken by giving vacuum expectation values (VEVs) along
$D$-- and $F$--flat directions to some of the 69 singlets $s^0_i$. We note that
cancellation of the Fayet--Iliopoulos $D$--term associated with an anomalous
$\U1$ requires that at least some of the VEVs be close to the string scale.
Choosing all the singlet VEVs of order the string scale, the $\U1$ gauge bosons
are decoupled, and we have
\begin{equation}
 G~\longrightarrow ~\SU{3}_c\times\SU{2}_\mathrm{L}\times \U1_Y   
 \times  G_\mathrm{hidden} \;,
\end{equation}
with $G_\mathrm{hidden}= \SO{6}\times\SU2$. This leads to complete separation
between the hidden and observable sectors.

One of the main problems of string models is the presence of exotic states at
low energies. Generically, such states are inconsistent with experimental data
and destroy gauge coupling unification. Even if the exotic states are
vector--like with respect to the SM, they are still harmful unless they attain
large masses. However, one cannot assign the mass terms at will. They must 
appear due to VEVs of some singlets and be consistent with string selection
rules [\onlinecite{Hamidi:1986vh},\onlinecite{Font:1988mm}]. In most cases, the
latter prohibit many of the required couplings such that the exotic states stay
light. One of the achievements of  our model is that all of the exotic
vector--like states can be given large masses consistently with string selection
rules.

To proceed, let us briefly summarize these rules \cite{Kobayashi:2004ud}. 
The coupling 
\begin{equation}
\Psi_1 \Psi_2\dots\Psi_n
\end{equation}
between the states $\Psi_i$ belonging to  twisted sectors $T_{k_i}$ (including
also  the untwisted sector)  to be allowed, the total twist has to be a multiple
of 6: $\sum_i k_i =0$ mod 6. There are further restrictions on the fixed points 
that can enter into this product, called a space group selection rule. For
example, in the $\SO4$ plane it amounts to $\sum_{i} (n^{(i)},n^{(i)'})=(0,0)  $
mod 2, where $(n^{(i)},n^{(i)'}) $ are the coordinates of the fixed points in
the orthonormal basis, $n^{(i)},n^{(i)'}=\{0,1\}$. Similar rules apply to the
$\SU3$ and the $\G2$ planes. Further, there is a requirement of gauge invariance
$ \sum_i p_i=0$, where $p_i$ are the (shifted) momenta in the $\E8 \times \E8$
gauge lattice. Finally, the $\boldsymbol{H}$--momentum  in  the compact 6D space
must also be conserved, $\sum_i R_1^{(i)}=-1$ mod 6, $\sum_i R_2^{(i)}=-1$ mod
3, $\sum_i R_3^{(i)}=-1$ mod 2, where $R_{1,2,3}$ are the
$\boldsymbol{H}$--momenta associated with the $\G2$, $\SU3$ and $\SO4$ planes,
respectively.

Based on these selection rules, we analyze allowed superpotential couplings
involving vector--like pairs of exotic fields $x_i \bar x_j$ and the SM singlets
$s_a$:
\begin{equation} 
 W= x_i \bar x_j \langle  s_a s_b \dots   \rangle \;.
\end{equation}
We find that all of the exotic states enter such superpotential couplings. This 
requires a product of up to 6 singlets.  In order to decouple the exotic states
one has to make sure that the corresponding mass matrices have a maximal rank
such that no massless  exotic states survive. An interesting feature of the
model is that there are exotic states with the SM quantum numbers of the
right--handed down quarks and lepton doublets. As a result these extra states
mix with those from the $\boldsymbol{16}$--plets of $\SO{10}$, such that the
observed matter fields are a mixture of both (cf.\
\cite{Asaka:2003iy},\cite{Casas:1988se}). In
particular, for down--type quarks we have the following mass matrix:
\[
\begin{array}{r|ccccccc}
 & \bar d_1  & \bar d_2  & \bar d_3  & \bar d_4  & \bar d_5  & \bar d_6 
  & \bar d_7 \\
\hline
d_1 & 
 s^5 & s^5 & s^5 & s^5 & s^5 & s^3 & s^3 \\
d_2 & 
 s^1 & s^1 & s^3 & s^3 & s^3 & s^3 & s^3 \\
d_3 &
 s^1 & s^1 & s^3 & s^3 & s^3 & s^3 & s^3 \\
d_4 &  
 s^6 & s^6 & s^6 & s^3 & s^3 & s^6 & s^6
\end{array}
\]
Here $s^n$ indicates that the coupling appears when a product of $n$ singlets is
included. Different entries with the same $n$ generally correspond to different
mass terms since they involve different singlets and Yukawa couplings.  This
mass matrix has rank 4 reflecting the fact that only 3 down--type quarks survive
and the others have  large masses, e.g. of the order of the string scale. Note
that higher $n$ does not necessarily imply suppression of the coupling: $\langle
s \rangle$ can be close to the string scale and, furthermore, the  combinatorial
coefficient in front of the coupling grows with $n$. 

An important phenomenological constraint on this texture comes from suppression
of $R$--parity violating interactions. It turns out that, in order to prohibit
$\bar u \bar d \bar d$ and similar couplings at the renormalizable level, the
$\bar d_{6,7}$ component in the massless $\bar d$--quarks must be suppressed.
This is achieved choosing appropriate directions in the space of singlet VEVs.

For the lepton/Higgs doublets  we have 
\[
\begin{array}{l|cccccccc}
 & \ell_1 & \ell_2 & \ell_3 & \ell_4 & \ell_5 & \ell_6 & \ell_7 & \ell_8\\
\hline 
\bar\ell_1 &
 s^3 & s^4 & s^4 & s^1 & s^1 & s^1 & s^1 & s^1 \\
\bar\ell_2 &
 s^1 &s^2 & s^2 & s^5 & s^5 & s^3 & s^3 & s^3 \\
\bar\ell_3 &
 s^1 &s^2 & s^2 & s^5 & s^5 & s^3 & s^3 & s^3 \\
\bar\ell_4 &
 s^1 &s^2 & s^2 & s^5 & s^5 & s^6 & s^3 & s^3 \\
\bar\ell_5 &
 s^1 &s^6 & s^6 & s^3 & s^3 & s^6 & s^3 & s^3
\end{array}
\]
This matrix has rank 5 which results in 3 massless  doublets of hypercharge
$-1/2$  at low energies. In order to get an extra pair of  (``Higgs'')  
doublets with hypercharge $-1/2$ and $1/2$,  one has to adjust the singlet VEVs
such that the rank reduces to 4. This unsatisfactory
fine tuning   constitutes the  well known  
supersymmetric $\mu$--problem.   A further constraint on the above texture comes
from the top Yukawa coupling: it is order one if the up--type Higgs doublet has
a significant component of  $\bar \ell_1$.

From the flavour physics perspective, it is interesting  that only the
right--handed  down--type quarks and the lepton/Higgs doublets mix with the
exotic states. Implications of  this phenomenon will
be studied elsewhere. Finally, the remaining  exotic states $m_i$ and
$s_i^{\pm}$  have full rank mass matrices and can be decoupled as well.

We have checked that the above decoupling is consistent with vanishing of the 
$D$--terms. This is implemented by constructing gauge invariant monomials out of
the singlets \cite{Buccella:1982nx} involved in the mass terms for the exotic
states. The $F$--flatness condition requires a more detailed study and will be
discussed elsewhere. Let us only mention that there are plenty of $F$--flat
directions in the field space,  for example, any direction in the 
39-dimensional  space of $T_2$,- $T_4$- and $U$-sector non--Abelian   singlets
is  $F$--flat to all orders   as long as the singlets from $T_{1,3}$  have zero
VEVs.   This is enforced by the $\boldsymbol{H}$--momentum selection rule for
the $\SO4$ plane. Whether the decoupling of all of the extra matter can be done
using exactly  flat directions or it requires isolated solutions to the
$F_i=D_a=0$  equations is currently under investigation. In any case, one can
show that  $F_i=D_a=0$  can be satisfied simultaneously on certain
low-dimensional manifolds in the field space, so the decoupling of extra matter
can be done consistently with supersymmetry.

The string selection rules have important implications for the matter Yukawa
couplings. In particular, in our setup the only large Yukawa coupling is that of
the top quark. The reason is that, at the renormalizable level, the types of
couplings allowed by the space group and the $\boldsymbol{H}$--momentum are
$UUU$, $T_1 T_2 T_3$, $T_1 T_1 T_4$, $U T_2 T_4$, $U T_3 T_3$. The third
generation up quark and quark doublet as well as the up--type Higgs doublet (up
to a mixing) are untwisted, so there is an allowed Yukawa interaction of the
type $UUU$ whose strength is given by the gauge coupling. The Yukawa couplings
involving the $T_3$ sector vanish since there is no SM matter in that sector.
The coupling $U T_2 T_4$ is incompatible with the $\SU2_\mathrm{L}\times\U1_Y$ 
symmetry, while the coupling $T_1 T_1 T_4$ is prohibited due to either gauge
invariance or decoupling of the exotic down--type quarks required by suppression
of $R$--parity violation. Therefore, all quarks and leptons apart from the top
quark are massless at the renormalizable level and their Yukawa interactions
appear due to higher order superpotential couplings. These are suppressed when
the involved singlets have VEVs below the string scale.

An important feature of the model is that it admits spontaneous supersymmetry
breakdown via gaugino condensation \cite{Ferrara:1982qs}. The $\SO6$ group of
the hidden sector is asymptotically free and its condensation scale depends on
the matter content. The $\boldsymbol{6}$--plets and the
$\boldsymbol{4}$,$\overline{\boldsymbol{4}}$--plets of $\SO6$ can be given large
masses consistently with the string selection rules. In this case, the
condensation scale is in the range $\sim 10^{11}-10^{13}\,\mathrm{GeV}$
depending on the  threshold corrections to the gauge couplings. Assuming that
the dilaton is fixed via the K\"ahler stabilization mechanism
\cite{Banks:1994sg}, gaugino condensation translates into supersymmetry breaking
by the dilaton.  The scale of the soft masses, $m_{3/2}$, depends on the details
of dilaton stabilization and can be in the TeV--range. 
   
\begin{figure}[t!]
\centerline{\CenterObject{\includegraphics{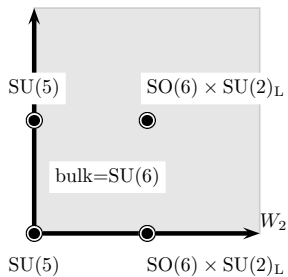}}
}
\caption{A 6D orbifold GUT limit with a large $\SO4$--plane compactification
radius.}
\label{fig3}
\end{figure}

Since our  model has no exotic states at low energies and admits TeV--scale soft
masses, it is consistent with gauge coupling unification. Then a natural
question to ask is what are the orbifold GUT limits
\cite{Kobayashi:2004ud,Forste:2004ie,Buchmuller:2004hv} of this model. That is,
what is the  effective field theory limit  in the energy range between the
compactification scale and the string scale  when some of the compactification
radii are significantly larger than the others. Such anisotropic
compactifications may mitigate the discrepancy between the GUT and string 
scales, and can be consistent with perturbativity for one or two large radii of
order $(2\times 10^{16} ~ {\rm GeV})^{-1}$
\cite{Witten:1996mz}. In our model, the intermediate orbifold picture  can have
any dimensionality between 5 and 10. For example, the 6D orbifold GUT limits are
(up to $\U1$ factors) :
\begin{eqnarray}   
 \SO4 ~\text{plane}& : &~\text{bulk GUT}=\SU6~, ~N=2 \;, \nonumber\\
 \SU3 ~\text{plane}& : &~\text{bulk GUT}=\SU8~,~N=2 \;, \nonumber\\
 \G2  ~\text{plane}& : &~\text{bulk GUT}=\SU6\times \SO4 ~,~N=4 \;, \nonumber
\end{eqnarray} 
where the plane with a ``large'' compactification radius is indicated  and $N$
denotes the amount of supersymmetry. In all of these cases, the bulk
$\beta$--functions of the SM gauge couplings coincide. This is because either
$G_\mathrm{SM}$ is contained in a simple gauge group or there is $N=4$
supersymmetry.  Thus, unification may occur below the string scale. 
(To check whether this is the case, logarithmic corrections from localized 
fields, contributions from vector--like heavy fields and string thresholds
have to be taken into account.)
The
SM gauge group is  obtained as an intersection of the gauge groups at the
different fixed points of the 6D orbifold (Fig.~\ref{fig3}).
%%%%%%%%%%%%%%%%%%%%%%%%%%%%%%%%%%%%%%%%%%%%%%%%%%%%%%%%%%%%%%%%%%%%%%%%
%%%%%%%%%%%%%%%%%%%%%%%%%%%%%%%%%%%%%%%%%%%%%%%%%%%%%%%%%%%%%%%%%%%%%%%%%%%

In this Letter, we have presented a heterotic string model which reproduces the
spectrum of the minimal supersymmetric SM  and is consistent with gauge coupling
unification.   The emerging picture has a simple geometrical interpretation. The
SM gauge group is obtained as an intersection of the local \E8 subgroups at
inequivalent orbifold fixed points. Two generations of quarks and leptons appear
as $\boldsymbol{16}$--plets localized at the fixed points with unbroken \SO{10}
symmetry, whereas the third `$\boldsymbol{16}$--plet' involves both bulk and
localized states. The Yukawa couplings do not exhibit \SO{10} relations.
The top quark Yukawa coupling is
related to the gauge couplings, while the other Yukawa couplings are due to
non--renormalizable interactions.   The model has a hidden sector which allows
for supersymmetry breaking via gaugino condensation.

Finally, let us remark that although this model is very  special, it is perhaps
not unique. In the $\Z6$-II orbifold with the $V_6$ shift given above, there are
roughly $10^4$ models (some of them may be equivalent) with the SM  gauge
group.  ${\cal O}(10^2)$  of them  have 3 matter generations  plus vector--like
exotic matter,  whereas we have so far  found only one model where the
vector--like matter can be decoupled  consistently with the string
selection rules. We plan to investigate this issue further.
It would also be interesting to understand the relation of this type of models
to other phenomenologically promising constructions \cite{Braun:2005ux}. %
%\\[0.05cm]

%\noindent
%{\bf Acknowledgements.} 
We thank T.~Kobayashi, H.~P.~Nilles and S.~Stieberger for valuable discussions. 
This work was partially  supported by the European Union 6th Framework Program
MRTN-CT-2004-503369 ``Quest for Unification'' and MRTN-CT-2004-005104
``ForcesUniverse''.

%%%%%%%%%%%%%%%%%%%%%%%%%%%%%%%%%%%%%%%%%%%%%%%%%%%%%%%%%%%%%%%%%%%%

\end{document}